\documentclass[runningheads,a4paper]{llncs}
\pdfoutput=1

\usepackage{url}
\usepackage{natbib}
\usepackage{amssymb}
\usepackage{graphicx}
\newcommand{\keywords}[1]{\par\addvspace\baselineskip
\noindent\keywordname\enspace\ignorespaces#1}

\begin{document}
\mainmatter  
\title{Maximum a posteriori estimation of piecewise arcs in tempo time-series}
\titlerunning{MAP estimation of piecewise arcs in tempo time-series}

\author{Dan Stowell\inst{1} and Elaine Chew\inst{1}}

\institute{Centre for Digital Music, Queen Mary, University of London\\ \email{dan.stowell@eecs.qmul.ac.uk}}

\maketitle

\begin{abstract}%
In musical performances with expressive tempo modulation, the tempo variation can be modelled as a sequence of tempo arcs.
Previous authors have used this idea to estimate series of piecewise arc segments from data.
In this paper we describe a probabilistic model for a time-series process of this nature,
and use this to perform inference of single- and multi-level arc processes from data.
We describe an efficient Viterbi-like process for MAP inference of arcs.
Our approach is score-agnostic, and together with efficient inference allows for 
online analysis of performances including improvisations, 
and can predict immediate future tempo trajectories.
\end{abstract}

\section{Introduction}
\label{sec:intro}

In various types of musical performance, one component of the musical expression is conveyed in the short-term manipulation of tempo, 
with tempo modulation reflecting musical phrase structure \citep{McAngusTodd:1992,Widmer:2003}.
This has motivated various authors to construct automatic analyses of the arc-shaped tempo modulations in recorded musical performances,
with or without score-derived information to supplement the analysis
\citep{McAngusTodd:1992,Widmer:2003,Chuan:2007}.
(See also \cite{Dannenberg:2011} who fit piecewise linear arcs to rock and jazz data, applying similar techniques but to genres in which the underlying tempo is held more fixed.)

Machine understanding of tempo, including its variability, can be useful in live human-machine interaction \citep{Dannenberg:1990,Robertson:2007}.
However most current online tempo-tracking systems converge to an estimate of the current tempo, modelling expressive variations as deviations rather than as components of an unfolding tempo expression.
In this paper we work towards the understanding of tempo arcs in a real-time system, paving the way for automatic accompaniment systems which follow the expressive tempo modulation of players in a more natural way.

We also consider tempo arcs within a probabilistic framework.
Previous authors have approached piecewise arc estimation using Dynamic Programming (DP) with cost functions based on squared error \citep{Chuan:2007,Dannenberg:2011}.
These are useful and can provide efficient estimation, but by setting the problem in a probabilistic framework (and providing the corresponding Viterbi-like DP estimator),
we gain some advantages:
prior beliefs about the length and shape of arcs can be expressed coherently as prior distributions;
measurement noise is explicitly modelled;
and the goodness-of-fit of models is represented meaningfully as posterior probabilities, which allows for model comparison as well as integration with other workflow components which can make use of estimates annotated with probability values.
Note that while we describe a fully probabilistic model, 
for efficient inference we will develop a Maximum A Posteriori (MAP) estimator, 
which returns only the maximum probability parameter settings given the priors and the data.

In the following we will describe our model of arcs in time-series data,
and develop an efficient MAP estimation technique based on least-squares optimisation and Viterbi-like DP.
The approach requires some kind of unsmoothed instantaneous tempo estimate as its input, which may come from a tempo tracker or from a simple measurement such as inter-onset interval (IOI).
We will then discuss how the estimator can be used for immediate-future tempo prediction, and how it can be applied to multiple levels simultaneously.
Finally we will apply the technique to tempo data from three professional piano performances, and discuss what the analysis reflects in the performances.

\section{Modelling and Estimation}
\label{sec:estimation}

For our basic model, we consider tempo to evolve as a function of metrical position (beat number) $x$ in a musical piece
as a series of connected arcs,
where each arc's duration, curvature and slope are independently drawn from prior distributions (to be described shortly).
Our model is deliberately simple, and agnostic of any score information that might be available.
To sample from this model, we pick an initial tempo at the starting time, then define a single upwards tempo arc which starts from that point,
and the tempo trajectory (speeding up and then slowing down) over a number of measures.
Any tempo data which may be measured during this interval is modelled as being drawn from the arc plus some amount of gaussian noise.
Once the ending breakpoint of this arc is reached, the next arc is sampled from the same priors, using the ending tempo as the new starting tempo.
Hence each tempo arc is conditionally independent of all previous observations once the starting tempo is determined, i.e.\ once the previous arc's parameters are fixed.
This assumption of conditional independence is slightly unrealistic, since it ignores long-range relationships between tempo arcs,
but it accounts for the most important interactions and makes inference tractable.

Our basic model is also only single-level, assuming that a single arc contributes to the current tempo at any moment,
rather than considering for example contributions from multiple timescales such as piece-level, movement-level, phrase-level and bar-level combined.
In Section \ref{sec:multiscale} we will consider a simple multi-scale extension of our technique, which we will apply in our analysis of piano performance data.
(For an alternative approach in which various components can be simultaneously active see \cite{McAngusTodd:1992}.)

\subsection{Fitting a Single Arc}

To fit a single arc shape to data, one can use standard quadratic regression, fitting a function of the form
\begin{equation}
f(x) = a + b x + c x^2,
	\label{eq:simplequad}
\end{equation}
and minimising the $L_2$ prediction error over the supplied data for $y \approx f(x)$.
In the Bayesian context, we wish to incorporate our prior beliefs about the regression parameters (here $a$, $b$ and $c$),
which is related to the optimisation concept of \textit{regularisation},
the class of techniques which aims to prevent overfitting by favouring certain parameter 
settings.
In fact, a gaussian prior on a regression parameter can be shown to be equivalent to the conventional $L_2$-norm regularisation of the parameters \citep[p.\ 153]{Bishop:2006},
summarised as:
\begin{equation}
	\textrm{regularisation coefficient} = \frac{\textrm{variance of gaussian noise}}{\textrm{variance of gaussian prior}}.
	\label{eq:equivalence}
\end{equation}
This equivalence is useful because it allows us to use common convex optimisation algorithms to perform the equivalent regularised least squares optimisation,
and they will yield the MAP estimate for the probabilistic model.

However, in this context a standard gaussian prior is not exactly what we require,
since we are expecting upwards arcs and not troughs -- we are expecting $c$ in Equation \ref{eq:simplequad} to be negative.
A more appropriate choice of prior might be a negative log-gaussian distribution, which allows us to specify a ``centre of mass'' for the arc shapes 
(expressed through the log-mean and log-standard-deviation parameters),
yet better represents our expectation that tempo arcs will always have negative curvature,
(almost) flat and extremely strongly curved arcs being equally rare.

The unconventional choice of prior might seem to remove the equivalence of the MAP regression technique with standard regularised least squres.
Yet if we rewrite our function to be
\begin{equation}
f(x) = a + b x - e^c x^2,
	\label{eq:modquad}
\end{equation}
then our prior belief about this modified parameter $c$ becomes a gaussian, yielding a negative-log-gaussian in combination with our function.
In addition, we will use a standard gaussian prior on $b$.
We could do the same for $a$ but instead we will use an improper uniform prior, for reasons which will be described in Section \ref{sec:multi}.
Therefore, our priors for Equation \ref{eq:modquad} will be gaussian priors on $b$ and $c$,
which can easily be converted to the equivalent $L_2$-regularisation terms for optimisation.

The strength of the regularisation (the value of the regularisation coefficient) reflects the specificity of our priors versus our data
-- specifically, the regularisation parameter is given by the noise variance divided by the prior variance \citep[p.\ 153]{Bishop:2006}.
Again, we see how the probabilistic setting helps to ground our problem, connecting the strength of the regularisation directly to our prior beliefs about the model and the data rather than manually-tuned parameters.

\subsection{Fitting Multiple Arcs}
\label{sec:multi}

\begin{figure*}[t]
	\centering
	\includegraphics [width=0.9\textwidth,clip,trim=20mm 10mm 20mm 10mm]  {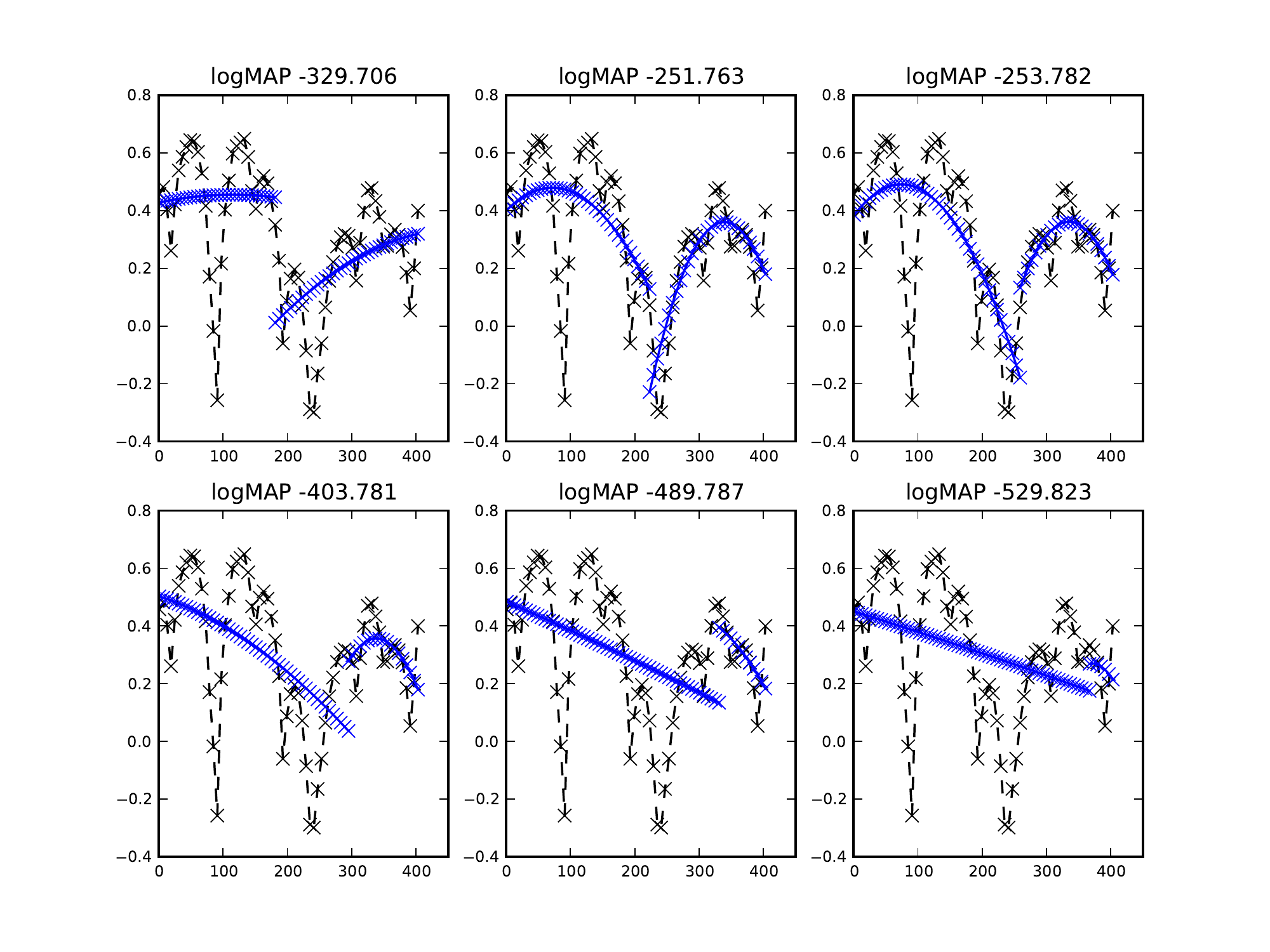}
	\caption{%
A selection of piecewise arc fits performed on a synthetic dataset.
For illustration purposes, we have manually specified a sequence of possible breakpoint locations,
and then performed a single-arc fit within each subsection. 
The ``logMAP'' (log of MAP probability) values quoted with each plot indicate the relative likelihood assigned to the depicted fit, given the prior parameters chosen. %
Prior parameters are the same for each of these plots.
The best-fitting plots have correspondingly higher (less negative) logMAP values.
}
\label{fig:toydatatwoarcs}
\end{figure*}

If a time-series is composed of multiple arcs and the breakpoints are known, 
then fitting multiple arcs is as simple as performing the above single-arc fit for each subsection of the time series (as in Figure~\ref{fig:toydatatwoarcs}).
Additionally, one should take care of the arc's dependence upon its predecessor (to enforce that they meet up), which is not shown in these plots.
In our case, we want to estimate the breakpoint locations as well as the arc shapes between those breakpoints.
This can be performed by iterating over all possible combinations of one breakpoint, two breakpoints, three (\ldots) for the dataset,
and choosing the result with the lowest cost (the highest posterior likelihood).

The Bayesian setting makes it possible to compare these different alternatives (e.g. one single arc vs.\ one arc for every datapoint)
without having to add arbitrary terms to counter overfitting;
instead, we specify a prior distribution over the arc durations, which in combination with the other priors and data likelihoods
yields a MAP probability for any proposed set of arcs.
In this paper we choose a log-normal prior distribution over arc durations.
See Figure \ref{fig:toydatatwoarcs} for some examples of different sets of arcs fitting to a synthetic dataset,
and the posterior (log-)probabilities associated.


In order for only a single tempo value to exist at each breakpoint 
(and not a discontinuous leap from one tempo to another),
we fit each arc under the constraint that its starting value equals the ending value of the previous arc.
This removes one degree of freedom from the function to be fit (Equation \ref{eq:modquad}) which otherwise has three free parameters.
We implement this by constraining the value of $a$ in the optimisation so that the function evaluates to the predetermined value at the appropriate time-point.
The least-squares optimisation therefore only operates on $b$ and $c$.

\subsection{Viterbi-like Algorithm}

The number of possible combinations of arcs for even a small time-series (such as Figure \ref{fig:toydatatwoarcs}) grows quickly very large,
and so it is impractical to iterate all combinations.
This is where Dynamic Programming (DP) can help.
Here we describe our DP algorithm, which, like the well-known Viterbi algorithm, 
maintains a record of the most likely route that leads to each of a set of possible states.
Rather than applying it to the states of a Hidden Markov Model, we apply it to the possibility that each incoming datum represents a breakpoint.

Assume that the first incoming datum is a breakpoint. 
(This assumption can be relaxed, in a similar way to the treatment of the final datum which we consider later.) 
Then, for each incoming datum ($x_n$, $y_n$), 
we find what would be the most likely path \textit{if it were certainly} a breakpoint.
We do this by finding the most appropriate past datum ($x_{n-k}$, $y_{n-k}$) which could begin an arc to the current datum -- 
where the appropriateness is judged from the MAP probability of said arc, 
combined with the MAP probability of the whole multiple-arc history that leads up to that past datum (recursively defined).

With our lognormal prior on the arc lengths (and with many common choices of prior), 
the probability mass is concentrated at an expected time-scale, and very long arcs are highly improbable \textit{a priori}.
Hence in practice we truncate the search over potential previous arc points to some maximum limit $K$ (i.e.\ $k \leq K$).

Thus, for every incoming data point we perform no more than $K$ single-arc fits, then store the details of the chosen arc, the MAP probability so far, 
and a pointer back to the datapoint at the start of the chosen single arc.
The simplest way to choose the overall MAP estimate is then to pick another definite breakpoint 
(for example, the last datum if the performance has finished)
and backtrack from there to recover the MAP arc path.

\subsubsection{Complexity}

The time complexity of the algorithm depends strongly on that of the convex optimisation used to perform a single-arc fit.
Assume that the complexity of a single-arc fit is proportional to the number of data points $k$ included in the fit, where $k \leq K$.
Then for each incoming data point a search is performed for one subset each of 2, 3, \ldots K data points,
which essentially yields an order $\mathcal{O}$($K^2$) process.
For online processing this is manageable if $K$ is not too large.
Analysing a whole dataset of $M$ points then has time complexity $\mathcal{O}$($K^2 M$).
(Compare this to the broadly similar complexity analysis of \cite{Dannenberg:2011}.)
The space complexity is simply $\mathcal{O}$($M$), 
or $\mathcal{O}$($K$) if the full arc history since the very beginning does not need to be stored.
This is because a small fixed amount of data is stored per datapoint.

\subsubsection{Predicting Immediate Future Arcs}

As discussed, if we know the performance has finished then we can find the Viterbi path leading to a breakpoint at the final data point received.
However, we would also like to determine the most likely set of arcs in cases where the performance might not have finished (e.g.\ for real-time interactive systems),
and thus where we do not wish to assert that the latest datum is a breakpoint.
We wish to be able to estimate an arc which may still be in progress.
If we can, this has a specific benefit of predicting the immediate future evolution of the tempo modulations (until the end of the present arc),
which may be particularly useful for real-time interaction.

We can carry this out in our current approach as follows.
Since an arc's duration (as well as the curve-fit) affects its MAP probability,
in the case where the latest arc may or may not be terminating we must iterate
over the arc's possible durations and pick the most likely.
To do this we choose a set of future time-points as candidate breakpoints, $x_{n+1}$ \ldots $x_{n+J}$ (e.g.\ an evenly-spaced tatum grid of $J=K$ future points).
Then we supply these data to the Viterbi update process exactly as is done with actual data,
but with no associated $y$ values.
These ``hypothetical'' Viterbi updates will use these time-points to determine the arc-lengths being estimated,
and in normalising the data subset,
but will not include them in the arc-fitting process.
It will therefore yield a MAP probability estimate for each of the time-points
as if an arc extended from the real data as far as this hypothetical breakpoint.
Out of these possibilities, the one with the highest MAP probability is the MAP estimate for an arc which includes the latest real datum and 
some portion of the hypothetical future points.
(The hypothetical Viterbi updates are not preserved: if more data comes in, it is appended to the Viterbi storage corresponding only to the actual data.)

\subsection{Multi-scale Estimation}
\label{sec:multiscale}

The model we describe operates at one level, with expected arc durations given by the corresponding prior.
Our model is adaptable to any time-scale by simply adapting the prior.
It does not however automatically lend itself to simultaneous consideration of multiple active timescales.

Multi-scale analysis can be carried out by analysing a dataset with one timescale,
then analysing the residual at a second timescale.
This residual-based decomposition has been used previously in the literature (e.g.\ \cite{Widmer:2003});
it requires a strong hierarchical assumption that the arcs at the first timescale do not depend at all on those at the second timescale,
while the second is subordinate to the first.
We consider this to be unrealistic, since there may well be interactions between the different timescales on which a performer's expression evolves.
However this assumption leads to a tractable analysis.

Note also that this approach to multi-scale estimation requires the first analysis to be completed (so that the residual is known)
before the second scale can be analysed.
Some DP approach may be possible to enable both to be calculated online,
but we have not developed that here.
For the present work, the single-scale Viterbi tracking is applicable and useful for online tracking,
while multi-scale analysis is an offline process, which we will next apply to modelling of pre-recorded tempo data.

\section{Analysis of Expressive Piano Performance}
\label{sec:piano}

We applied our analysis to an existing set of annotations of three performances of Beethoven's \textit{Moonlight Sonata}.
The annotations by Elaine Chew have previously been analysed by Chew with reference to observations noted by Jeanne Bamberger \citep{Chew:2012}.
For each of three well-known performances of the piece---by Daniel Barenboim (1987), Maurizio Pollini (1992) and Artur Schnabel (2009)---%
the first 15 bars have been annotated with note onset times, which correspond to regular triplet eighth-note timings.

We implemented the algorithm in Python, using the \texttt{scipy.optimize.fmin} optimiser to solve individual regressions. 
Source code is available.%
\footnote{\url{https://code.soundsoftware.ac.uk/projects/arcsml}}
(Note that this development implementation is not generally fast enough for real-time use.)

Instantaneous tempo was derived from these inter-onset intervals, then analysed using a two-pass version of our algorithm:
first the data was analysed using an arc-duration prior centred on four bars;
then the residual was analysed using an arc-duration prior centred on one bar.
This choice of timescales is a relatively generic choice which might reasonably be considered to reflect a performer's short-term and medium-term state;
however it might also be said to be a form of basic contextual information about the relevant timescales in the current piece.
For the current study, we confine ourselves to priors with log-normal shapes,
though an explicitly score-derived or corpus-derived prior could have a more tailored and perhaps multimodal shape.

\begin{figure*}[t]
	\centering
	\includegraphics [width=0.8\textwidth,height=0.28\textheight,clip,trim=10mm 2mm 10mm 6mm]  {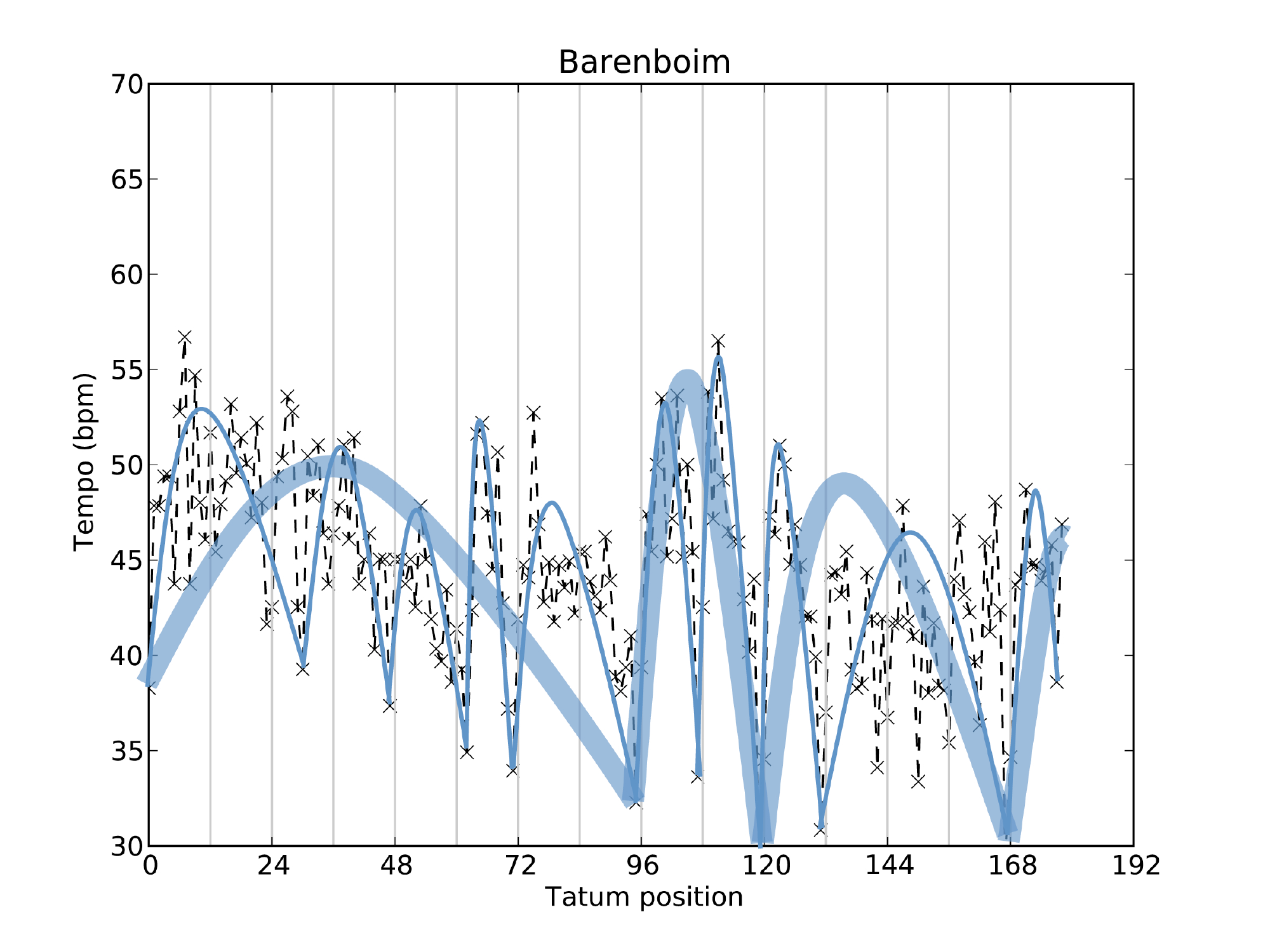} 
	\includegraphics [width=0.8\textwidth,height=0.28\textheight,clip,trim=10mm 2mm 10mm 6mm]  {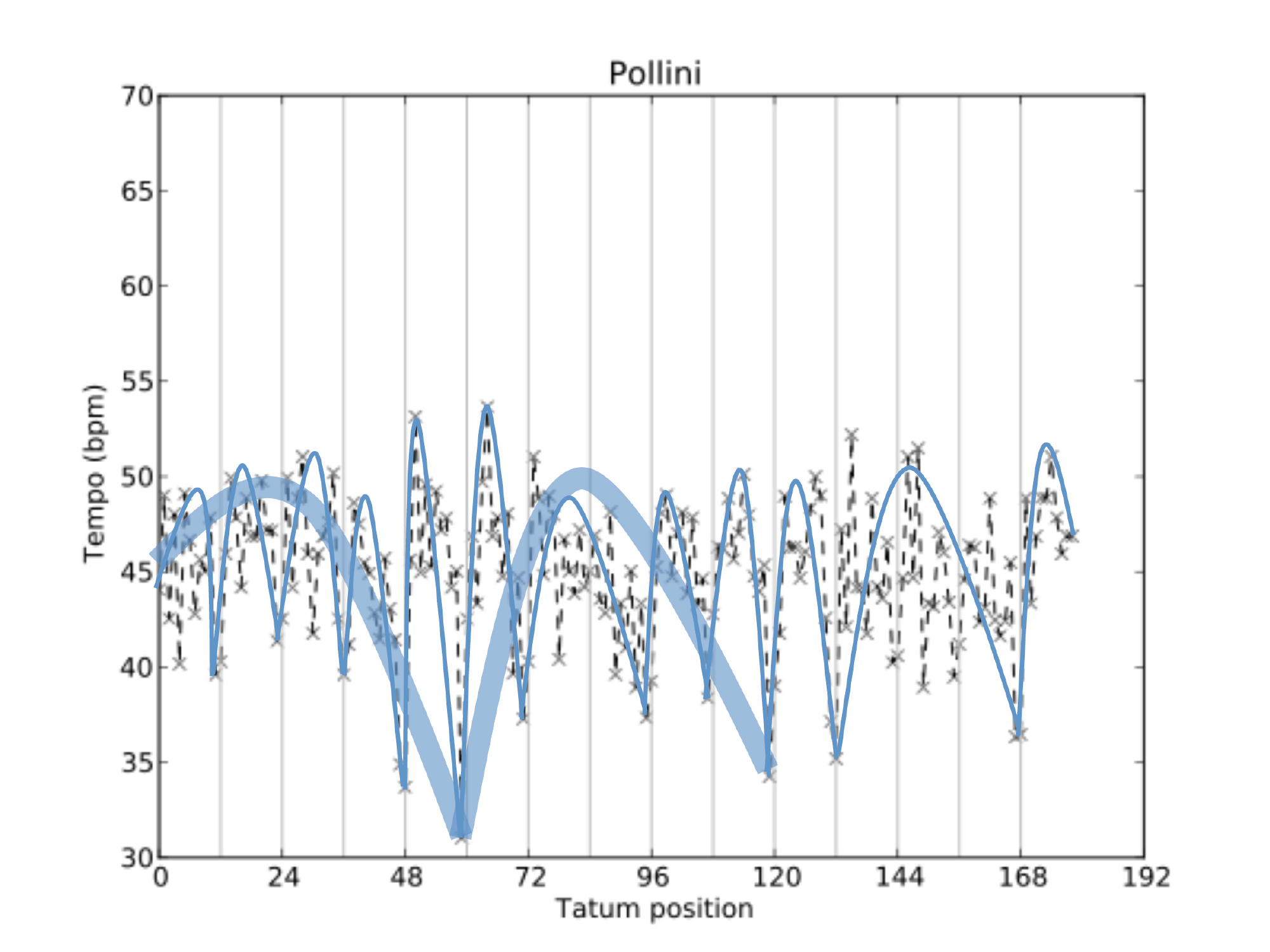} 
	\includegraphics [width=0.8\textwidth,height=0.28\textheight,clip,trim=10mm 2mm 10mm 6mm]  {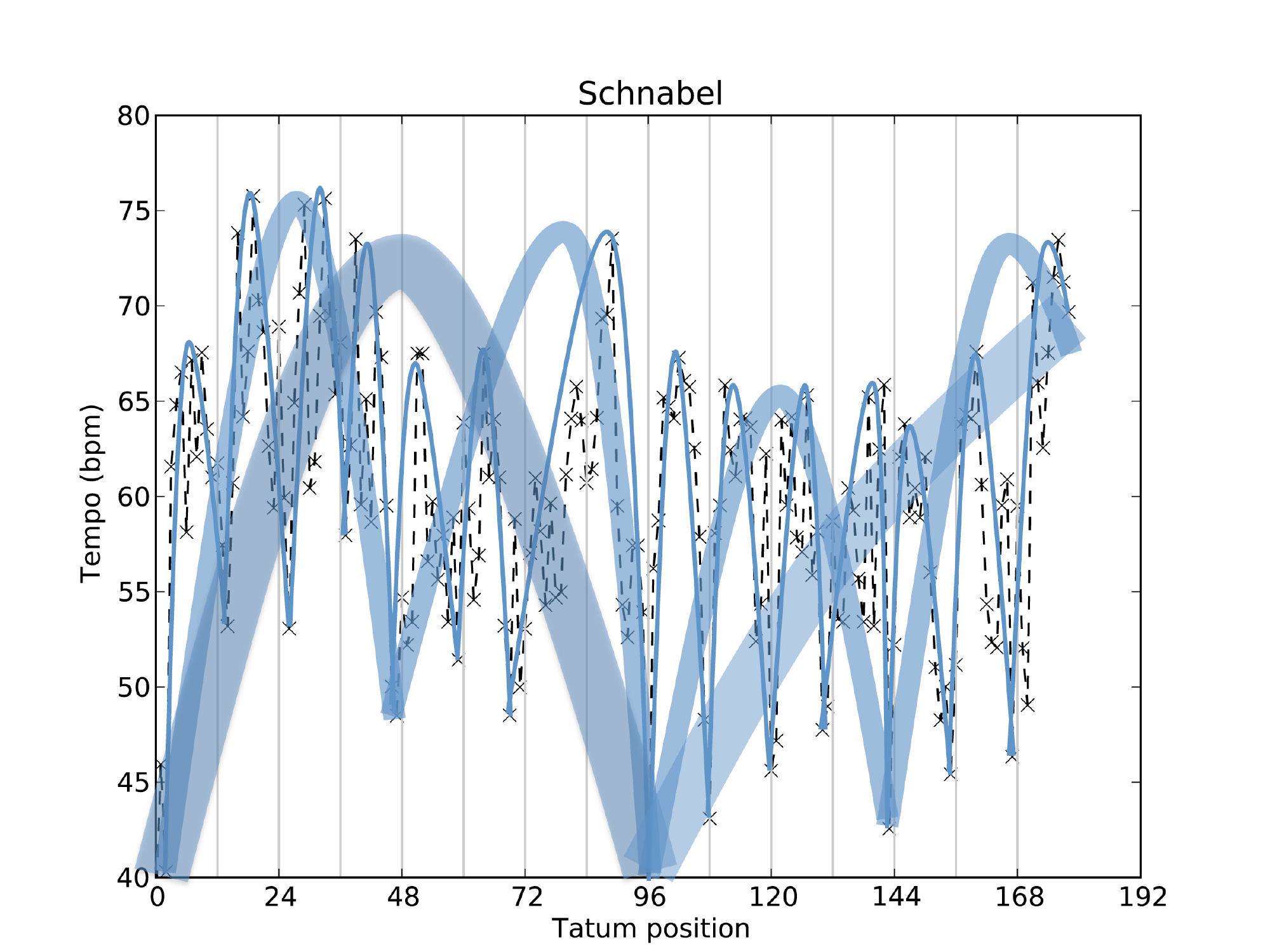}
	\caption{%
Manual analyses of performances by each of three pianists (Barenboim, Pollini, Schnabel).
Two to three levels of arcs are drawn by visual inspection for each tempo time series.
No higher level arcs are drawn when it is uncertain that one exists.
}
\label{fig:manual_analyses}
\end{figure*}

\begin{figure*}[t]
	\centering
	\includegraphics [width=0.8\textwidth,height=0.28\textheight,clip,trim=10mm 2mm 10mm 6mm]  {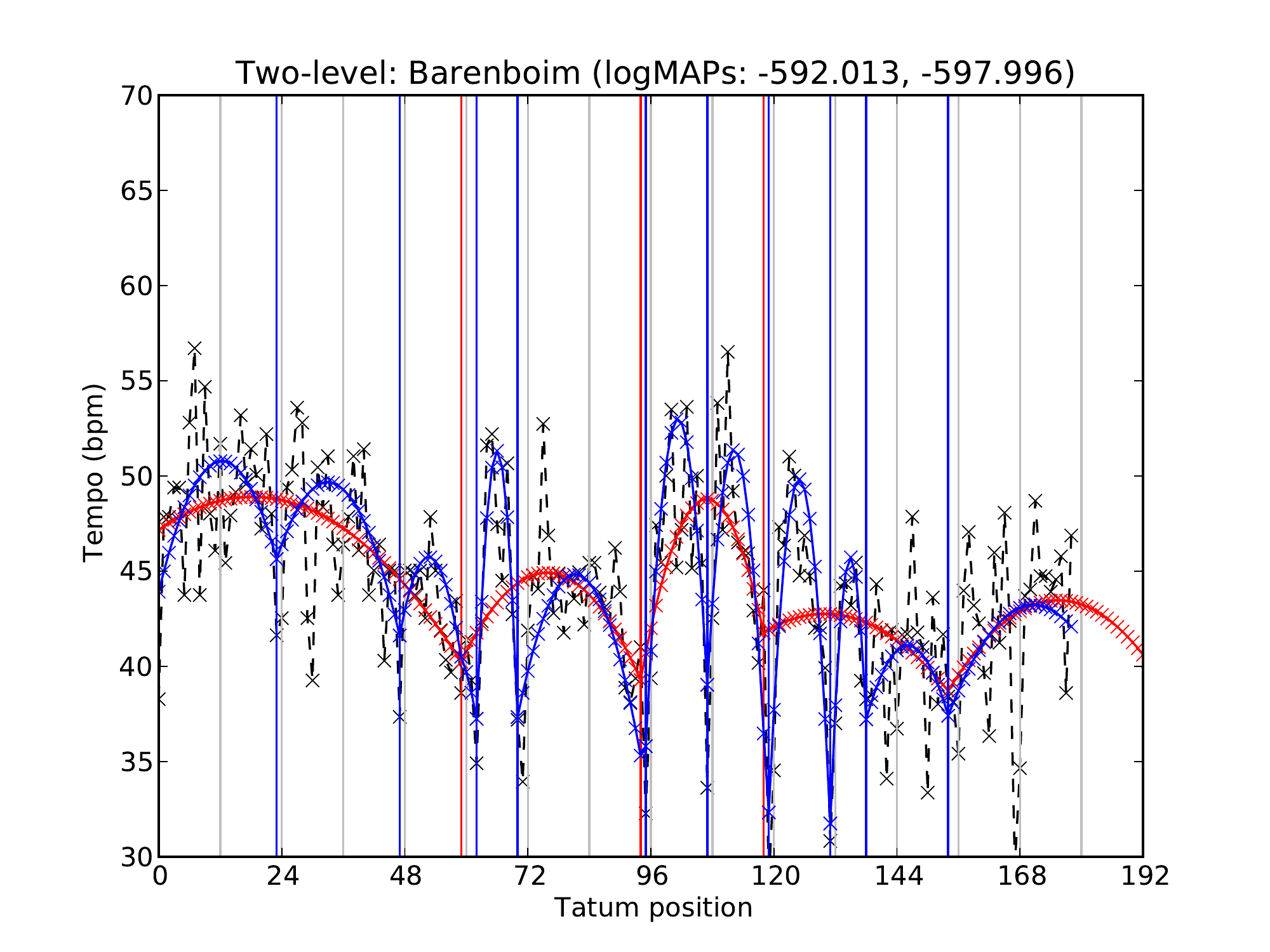} 
	\includegraphics [width=0.8\textwidth,height=0.28\textheight,clip,trim=10mm 2mm 10mm 6mm]  {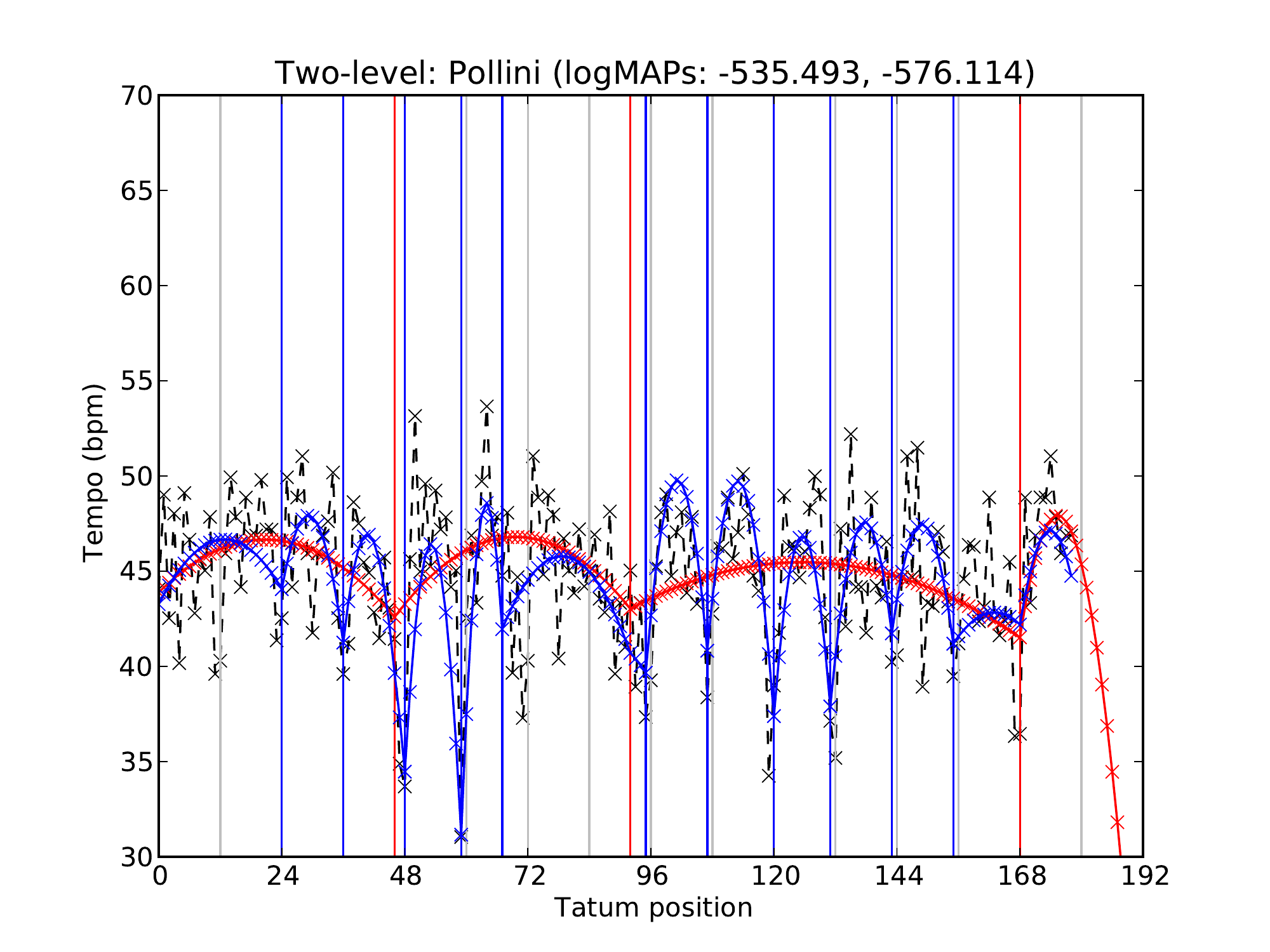} 
	\includegraphics [width=0.8\textwidth,height=0.28\textheight,clip,trim=10mm 2mm 10mm 6mm]  {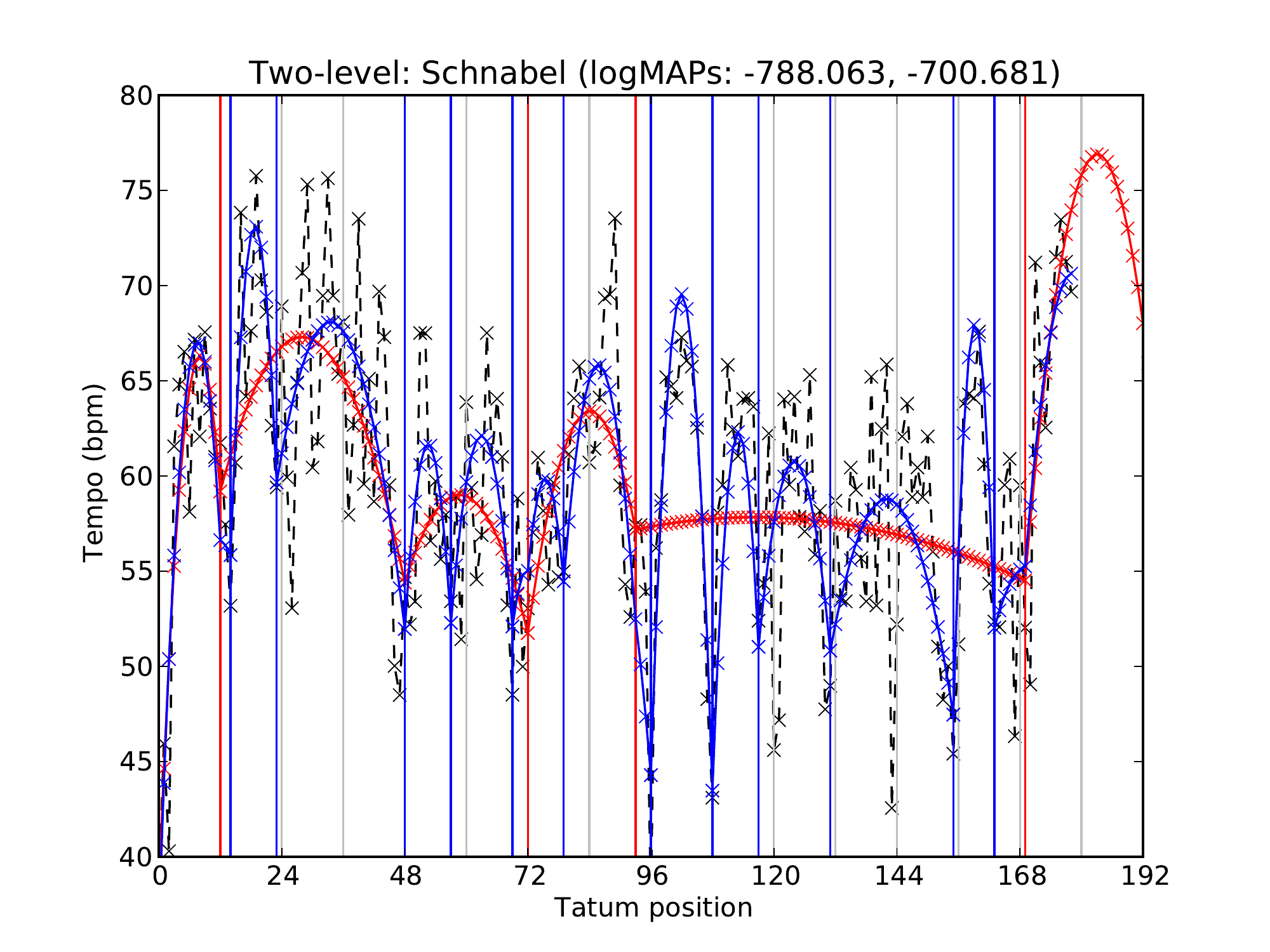}
	\caption{%
Two-level analysis of performances by each of three pianists (Barenboim, Pollini, Schnabel). %
In each plot, the first long-scale fit (centred on the four-bar timescale) is depicted in red,
and the second shorter-scale fit (centred on the one-bar timescale) is given in blue.
The second fit is pre-offset by the first, meaning the blue arcs display the combined model produced by both timescales combined.
Annotated data finish at tatum 180; where the MAP choice extends beyond that, we show the predicted immediate future arc.
}
\label{fig:twoarcsthree}
\end{figure*}

\begin{figure*}[t]
	\centering
	\includegraphics [width=0.8\textwidth,height=0.28\textheight,clip,trim=10mm 2mm 10mm 6mm]  {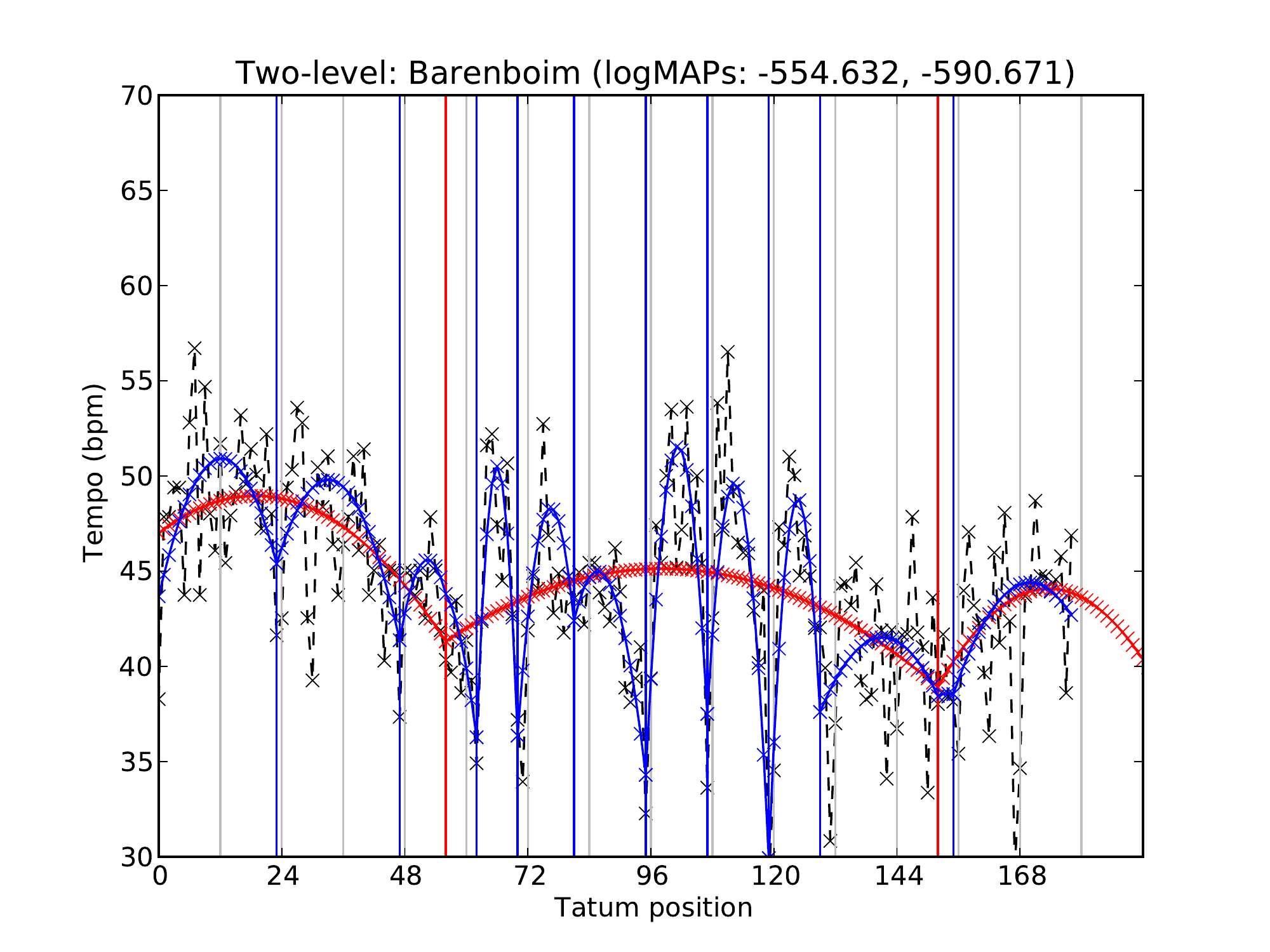} 
	\includegraphics [width=0.8\textwidth,height=0.28\textheight,clip,trim=10mm 2mm 10mm 6mm]  {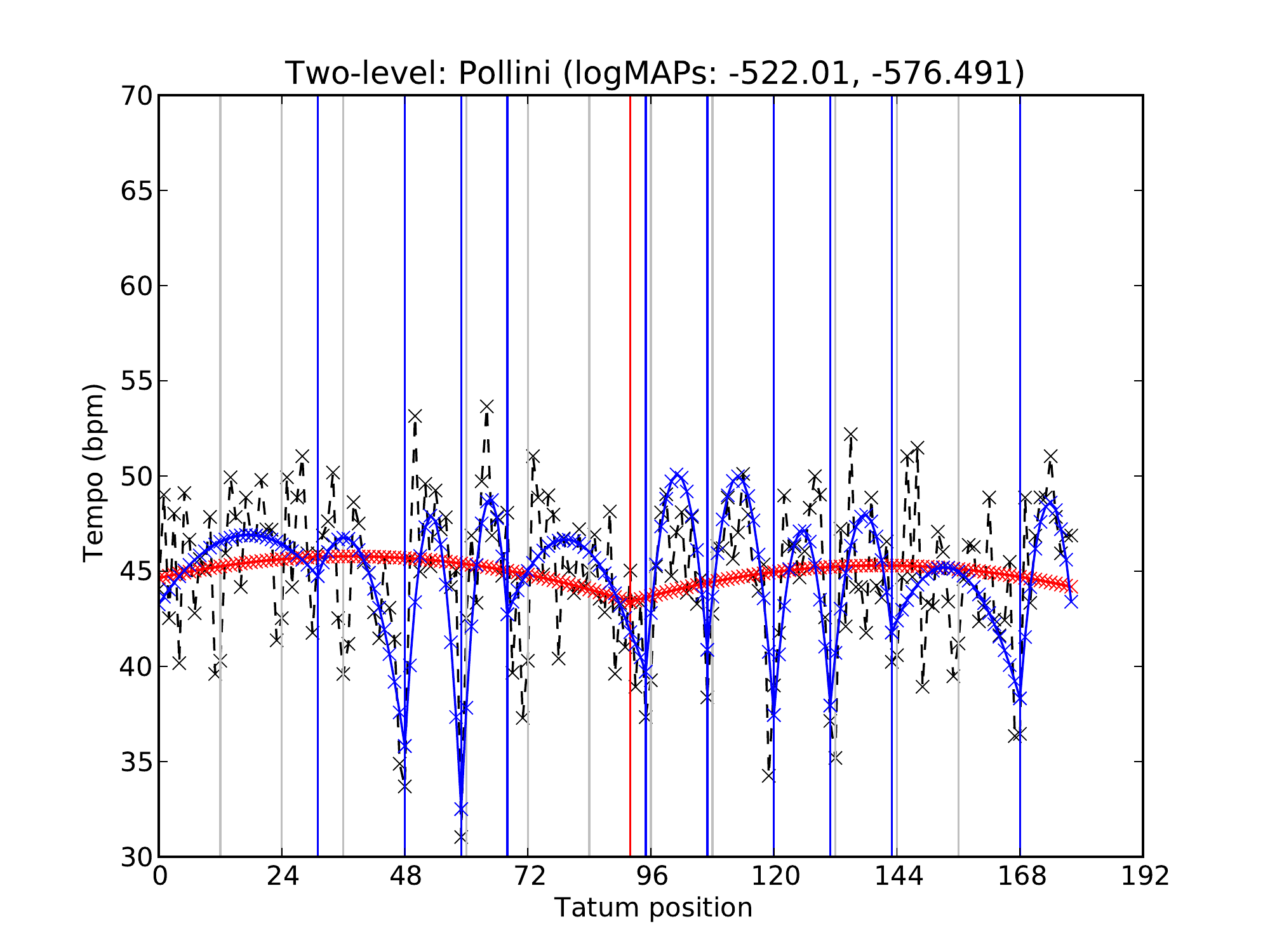} 
	\includegraphics [width=0.8\textwidth,height=0.28\textheight,clip,trim=10mm 2mm 10mm 6mm]  {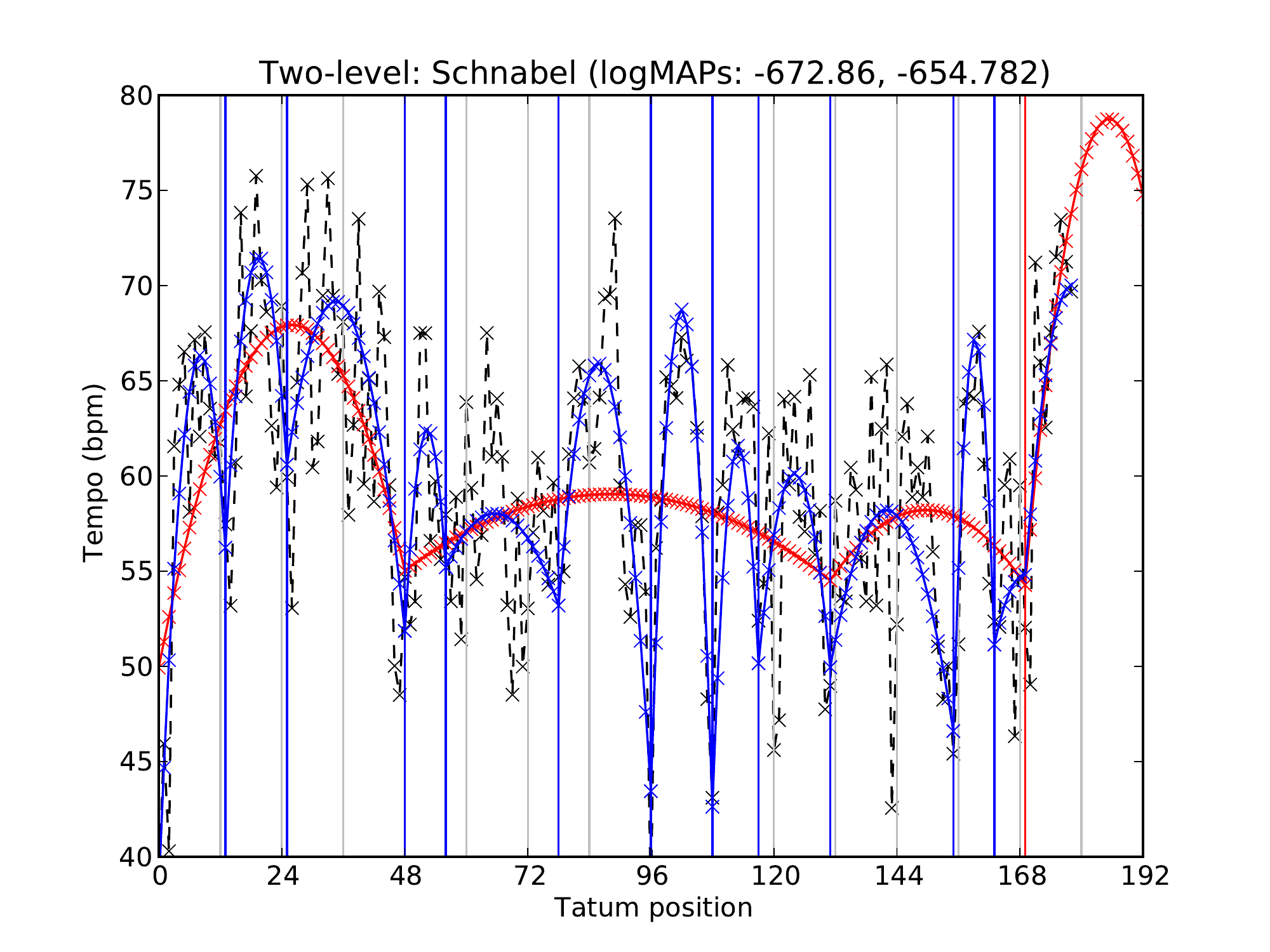}
	\caption{%
As Figure \ref{fig:twoarcsthree} but with the standard deviation of the noise prior set at 4.0 rather than 3.0.
}
\label{fig:twoarcsthreeB}
\end{figure*}

\begin{figure*}[t]
	\centering
	\includegraphics [width=1.0\textwidth,height=0.28\textheight]  {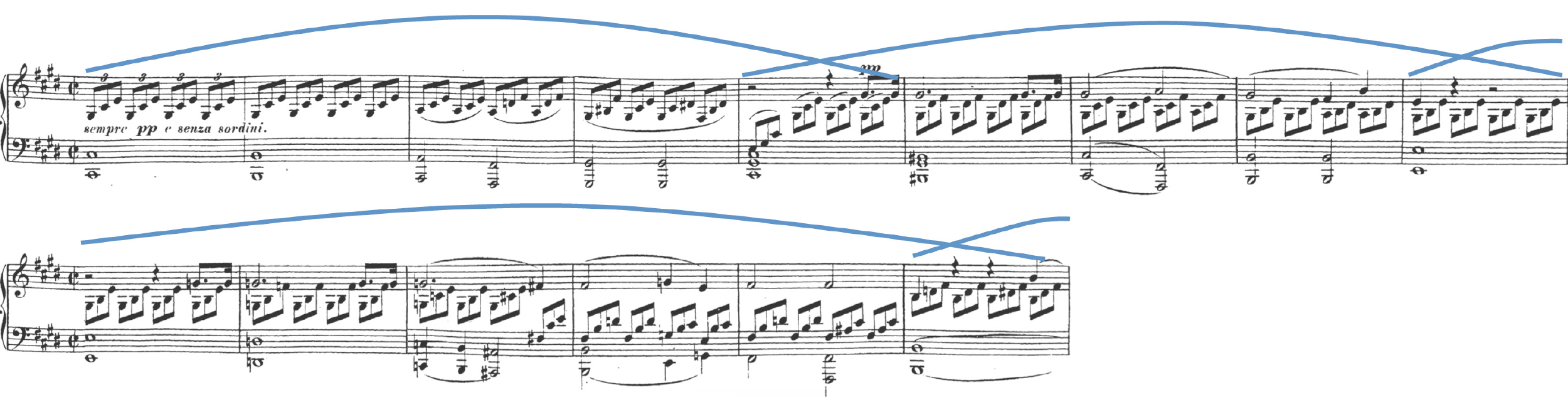}
	\caption{%
Possible set of overlapping phrases in the first 15 bars of the Moonlight Sonata.
}
\label{fig:moonlight_phrases}
\end{figure*}

Figure \ref{fig:manual_analyses} shows a set of manual annotations of hierarchically embedded phrases,
and Figure \ref{fig:twoarcsthree} shows the automatically computed results.
The automatic analyses show some notable similarities with the manual one at the shorter time scale,
and significant differences at the longer time scale.
The difficulty of the longer time scale analysis may be explained by Figure \ref{fig:moonlight_phrases}, which shows one plausible set of phrase groupings for this excerpt;
the overlapping \{5,5,7\} bar phrases do not fit easily into a four-bar duration framework.

Nevertheless, the longer time scale analysis (centred on four bars) highlights differences between the performances:
Pollini's performance appears to contain relatively little variation on this level, as the fit yields long and shallow arcs,
with breakpoints near positions 48, 96 and 168 (structurally important positions; 96 is where the key-change occurs).
On the other hand, both Barenboim and Schnabel's tempo curves exhibit fairly deep and varied arcs.
Schnabel's performance exhibits the most dramatic variation in the first four bars until around measure 48:
this first four-bar section corresponds to the opening statement of the basic progression, before the melody enters in the fifth bar (and the underlying progression repeats).
Bamberger described Schnabel as performing them ``as if in one long breath'' (quoted in \cite{Chew:2012}), not quite reflected in our automatic analysis.

On the shorter time scale, the analysis tends to group phrases into one-bar or two-bar arcs.
Aspects of the musical structure are reflected in the arcs observed.
Sections of the melody which lend themselves to two-bar phrasing (e.g.\ 72--96) are generally reflected in longer arcs crossing bar lines.
Conversely, in the region 96--132 the change to the new key unfolds as each new chord enters at the start of a bar,
and the tempo curves for all three performers reflect an expressive focus on this feature, with one-bar arcs which are more
closely locked to the bar-lines than elsewhere.
Note that in this section Schnabel matches Pollini in exhibiting a long and shallow arc on the slow timescale, with all the expressive variation concentrated on the one-bar arcs.

Over the excerpt generally, the breakpoints for Schnabel are further away from the barline than the others, as was observed in Chew's manual analysis.
We can quantify this by measuring the mean deviances of arc endpoints from the barlines in each performance.
The resulting mean deviances confirm our observations (Table \ref{tbl:dev}).

\begin{table}[t]
	\centering
	\begin{tabular}{l | r}
	Performer & Mean deviance (bars) \\
		\hline
	Barenboim & 13.9\% \\
	Pollini &    8.3\% \\
	Schnabel &  15.5\% \\
	\end{tabular}
\caption{Mean deviance from the barlines of the arc endpoints inferred for each performance, averaged over the short-timescale arcs in each case.}
\label{tbl:dev}
\end{table}

We have extended the plots slightly beyond the 180 annotated data points, to illustrate the immediate-future predictions made by the model.
(This is done for both timescales, though only the longer timescale (in red) shows noticeable extended arcs.)
All the performers, and especially Schnabel, exhibit an acceleration towards the end of the annotated data, reflected in the predictions of an upward arc followed by a gradual slowing over the next bar.
This type of prediction is plausible for such expressively-timed music.

To illustrate the effect that the prior parameters have upon the regression, 
Figure \ref{fig:twoarcsthreeB} shows the same analysis
as Figure \ref{fig:twoarcsthree} but with the standard deviation of the noise prior set at 4.0 rather than 3.0.
The increase in the assumed noise variance leads the algorithm to ``trust'' the data less and the prior slightly more (cf.\ Equation \ref{eq:equivalence}).
In our example, some of the breakpoints for the long-term arcs (in red) have changed, losing some detail, 
though most of the detail of the second-level analysis (in blue) is consistent.

\section{Conclusions}
\label{sec:conc}

We have described a model with similarities to some previous piecewise-arc models of musical expression,
but with a Bayesian formulation which facilitates model comparison and the principled incorporation of prior beliefs.
We have also described an efficient Viterbi-like Dynamic Programming approach to estimation of the model from data.
The approach provides scope to apply the model to real-time score-free performance tracking,
including prediction of immediate future tempo modulation.
Source code for the algorithm (in Python) is available.

We have applied the model in a two-level analysis to data from expressive piano performance,
illustrating the algorithm's capacity to operate at different time-scales, 
and to recover expressive arc information that corresponds with some musicological observations regarding phrasing and timing.

Further research would be needed to develop a model of multiple simultaneously-active levels of expression which can be applied online
as with our single-level Viterbi-like algorithm.
Similar arcs have been observed and analysed in loudness information extracted from performances \citep{Cheng:2008}.
It would also be useful to combine loudness information with tempo information in this model.



\end{document}